\documentclass[pdftex,12pt]{article}
\pdfoutput=1
\usepackage{natbib}
\usepackage[pdftex]{graphicx}        
\usepackage{color}           
\usepackage{url}             
\usepackage[breaklinks=true]{hyperref}

\usepackage{mbenotes}

\newcommand{\etal}{{\it et al.}}

\title{\vspace{-2cm}Direct Measurement Results of the Time Lag of LOS-Velocity
Oscillations Between Two Heights in Solar Faculae and Sunspots}

\author{N.~Kobanov, D.~Kolobov, A.~Kustov, S.~Chupin, \\
    and A.~Chelpanov \\
    \small{Institute of Solar-Terrestrial Physics} \\
    \small {of Siberian Branch of Russian Academy of Sciences, Irkutsk, Russia} \\
    \small {email: \url{kolobov@iszf.irk.ru}}
    }
\date{\small{[\href{http://dx.doi.org/10.1007/s11207-013-0247-2}
{This article was firstly published in \textit{Solar Physics}}]}}

\begin{document}
\maketitle

\begin{abstract}
We present an investigation of line-of-sight (LOS) velocity oscillations in
solar faculae and sunspots. To study the phase relations between
chromospheric and photospheric oscillations of the LOS velocity, we measured the time
lag of the chromospheric signal relative to the photospheric one for several
faculae and sunspots in a set of spectral line pairs. The measured time lags are
different for different objects. The mean measured delay between the
oscillations in the five-minute band in faculae is 50\,s for the Si\,{\sc i}
10\,827\,\AA\,--\,He\,{\sc i} 10\,830\,\AA\ pair; for the pair Fe\,{\sc
i} 6569\,\AA\,--\,H$\alpha$ 6563\,\AA\ the mean delay is 20\,s; for the pair
Fe\,{\sc i} 4551\,\AA\,--\,Ba\,{\sc ii} 4554\,\AA\ the mean delay is 7\,s; for
the pair Si\,{\sc i} 8536\,\AA\,--\,Ca\,{\sc ii} 8542\,\AA\ the mean delay is
20\,s. For the oscillations in the three-minute band in sunspot umbrae the mean
delay is 55\,s for the Si\,{\sc i} 10\,827\,\AA\,--\,He\,{\sc i} 10\,830\,\AA\
pair; for the Fe\,{\sc i} 6569\,\AA\,--\,H$\alpha$ 6563\,\AA\ pair it was not
possible to determine the delay; for the Fe\,{\sc i} 4551\,\AA\,--\,Ba\,{\sc ii}
4554\,\AA\ pair the mean delay is 6\,s; for the Si\,{\sc i}
8536\,\AA\,--\,Ca\,{\sc ii} 8542\,\AA\ pair the mean delay is 21\,s. Measured
delays correspond to the wave propagation speed which significantly exceeds the
generally adopted speed of sound in the photosphere. This raises the question
of the origin of these oscillations. The possibility that
we deal with slow MHD waves is not ruled out.

\end{abstract}

\section{Introduction}
     \label{sec:intro}
\par Solar faculae and sunspots can play an important role in the energy exchange
among different layers of the solar atmosphere. The oscillatory behaviour of the
chromosphere in these regions is a well-established phenomenon. The first
investigations into wave propagation, including studies of waves propagating in
a vertical direction from the photosphere to the chromosphere, are date from the
early 1970s
\citep{beckers1969uf,giovanelli1972,zirin1972,moore1975,giovanelli1978a,lites1979,gurman1982,lites1986b,gurman1987a}.
But the interest in this theme is still considerable today
\citep{georgakilas2000a,christopoulou2000,rouppe2003,balthasar2003,kobanov2006,nagashima2007,bloomfield2007a,balthasar2010,rez2012ApJ,sych2012}.
The most reliable evidence of upward propagating waves is to directly measure
time delays between the photospheric and chromospheric Doppler velocity
oscillations. Therefore, three-minute oscillations in sunspots and five-minute oscillations
the faculae are particularly interesting. The main goal of this article is to
directly measure the time delay for various pairs of spectral lines.

\par We selected measurements with long time series, high cadence, a good
signal-to-noise ratio, and a suitable object location. The latter implies, for
example, that data sets were taken close to the solar disc centre to avoid
projection effects. It is reasonable to observe sunspots of regular form to
facilitate the interpretation and data reduction. The choice of spectral line
pairs is very importance as well. For reliability, the analysis is to cover at
least several objects.

\section{Observations and Data}
     \label{sec:obsdata}

\par The observational spectral data analysed here were obtained with the
\textit{Horizontal Solar Telescope} at the Sayan Solar Observatory
\citep{kobanov09}. The telescope is elevated 6\,m above the ground, is windproof,
and equipped with a dedicated system to suppress atmospheric turbulence
\citep{ham1973}. The instrument mirrors are of 80~cm diameter. The focal length
of the main mirror is 20\,m, hence the possible spatial resolution is 0.2$''$.
However, due to the effects of the Earth's atmosphere, the real resolution is
about 1$''$\,--\,1.5$''$. The guiding system performs targeting and object
capturing with a precision of at least 1$''$, and compensates for the
rotation of the Sun -- the target is fixed on the spectrograph's slit during the
experiment. The image jitter on the slit depends on the observational conditions
and can reach 2$''$ on average in the series we investigated here. Some of the time series
were accompanied by an overview scan of the area under investigation. The
focal length of the spectrograph is 7\,m, and the diffraction grating parameters
are 200$\times$300\,mm, 600 grooves per mm.

\par We made observations in pairs of the following lines: Si\,{\sc i}
10\,827\,\AA\ and He\,{\sc i} 10\,830\,\AA; Si\,{\sc i} 8536\,\AA\ and Ca\,{\sc
ii} 8542\,\AA; Fe\,{\sc i} 6569\,\AA\ and H$\alpha$ 6563\,\AA; Fe\,{\sc i}
4551\,\AA\ and Ba\,{\sc ii} 4554\,\AA. The data contain 14 time series
in faculae and 17 time series in sunspots (see
Tables~\ref{tbl:faculae} and \ref{tbl:spots}). Two digital cameras were used in
the observations:  Princeton Instruments (PI) RTE/CCD 256H (256$\times$1024,
visible -- near infrared range), and FLIR SC-2200 (256$\times$320, infrared
range). One PI camera pixel corresponds to 0.23$''$, and the FLIR camera pixel
corresponds to 0.3$''$. The spectral resolution of the PI camera for the Fe\,{\sc i}
6569\,\AA\ and H$\alpha$ 6563\,\AA\ lines is 8\,m\AA\ per pixel; for Si\,{\sc i}
10\,827\,\AA\,--\,He\,{\sc i} 10\,830\,\AA\ it is 20\,m\AA\ per pixel. The
resolution of the FLIR camera is 25\,m\AA\ per pixel for the Si\,{\sc i}
10\,827\,\AA\,--\,He\,{\sc i} 10\,830\,\AA\ pair.

\par For three time series (Table~\ref{tbl:spots}, Nos. 15\,--\,17) the following
polarisation optics were used: FLC modulator (DisplayTech), quarter-wavelength
plate and Nicol prism. This allows one to register two frames $I+V$, $I-V$ for
the whole spectrum domain that is captured by the FLIR camera. The optical
scheme is described in \cite{kolobov2008}. The single-frame exposure
time is 39\,ms (25\,Hz cadence). When using two frames obtained in 80\,ms no
object displacement along the slit was revealed. Ten even and odd frames, with
respect to the modulation state, were averaged to form two frames ($I+V$, $I-V$)
with a 0.8\,s total acquisition time. For the other time series, taken without
polarisation optics, the cadence varies from 0.5 to 10\,s, and the total
duration ranges from 42 to 198 minutes.

\par Processing spectrograms included standard procedures: subtraction of dark
frame and flat-field correction. We obtained line-of-sight (LOS) velocity signals
via different techniques: \textit{i}) the Doppler compensation technique or, in
other words, the lambda-meter \citep{severny58,rayrole1967}, and
\textit{ii})~tracking of the Stokes-V profile zero-crossing position. In the
lambda-meter technique, two virtual slits were used to determine the spectral
position of a chosen line. The distance between the slits was set and
remained constant during the measurements. Initially, the slits were placed at
equal distances from the line centre, and hence the measured intensities are
equal. If in the next spectrogram the line shifted, the intensities changed.
By displacing the slits to make their intensities equal, one can
determine the new location of the line. The Stokes-V zero-crossing was
calculated as a mean value of the positions of the red and blue lobes for a
chosen spectral line. The Stokes-V data are available only for sunspots series
Nos.\,15\,--\,17 (Table~\ref{tbl:spots}).

\par To analyse oscillations of a specific frequency, we used band
filtration with a sixth-order Morlet wavelet. The filtration frequency range
for both the photospheric and chromospheric signals was selected based on the
fast Fourier transform (FFT) power spectra, obtained for different positions along the slit. We chose the
position so that both signals contained peaks in the same frequency band. We used
the algorithm described by \cite{torrence1998} to derive the FFT power spectra
($1/\sigma^2$ normalisation, where $\sigma^2$ is the variance of a time series),
to wavelet-filter the daya and estimate the statistical significance. The
power of the chromospheric and photospheric five-minute oscillations of the LOS velocity in
faculae significantly exceeds the noise level. Oscillations in this band reveal
themselves clearly in the original (unfiltered) signals. In practice, one does not need
to check the statistical significance of the FFT power at these
periods. The same is true for the chromospheric three-minute band oscillations in
sunspot umbrae of the time series. Estimating the statistical
significance was useful for the photospheric umbral oscillations, which are
suppressed relative to the surrounding regions and might have amplitudes that
are below or around the noise level. The series demonstrated a corresponding FFT
power at about the 3$\sigma^2$ level. Here it was not possible to obtain
an unambiguous phase delay between the photosphere and the chromosphere.

\par
\begin{figure} 
\centerline{
\includegraphics[width=11.5cm]{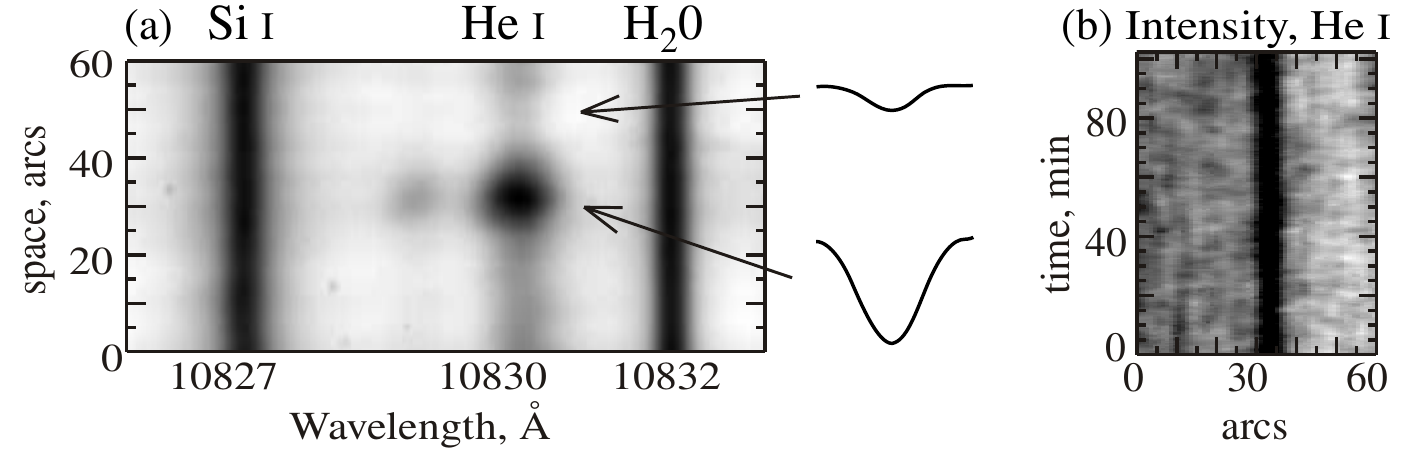}
}
\caption{Time series No.\,8. Overview of the data. Panel (a): spectrum
for He\,{\sc i} 10\,830\,{\AA} demonstrating the increase in line depth in
faculae; panel (b): He\,{\sc i} 10\,830\,\AA\ line intensity.
        }
\label{fig:fac-data-overview}
\end{figure}

\begin{figure} 
\centerline{
\includegraphics[width=5cm]{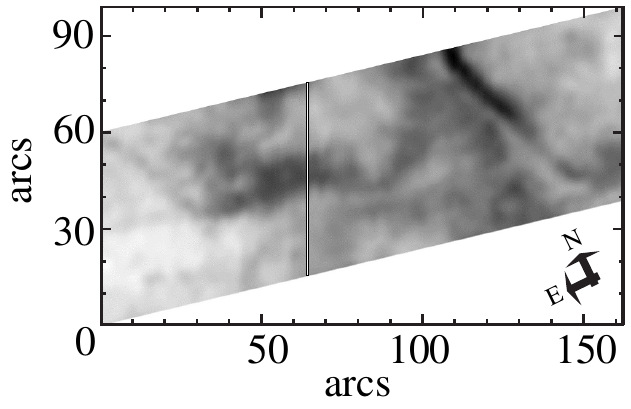}
}
\caption{Spatial scan for the faculae (time series No.\,8).
The vertical line marks the slit position.
        }
\label{fig:fac-scan}
\end{figure}

 The He\,{\sc i} 10\,830\,\AA\ line profile depth in faculae sharply
increases (Figure~\ref{fig:fac-data-overview}\,(a)) compared to the surrounding
area. According to our observations, the increase in different faculae varies
from 2 to 5. We used this feature of the He\,{\sc i} 10\,830\,\AA\ line to
precisely point at faculae near the disc centre. At first, we chose an object
from full-disc images in the Ca\,{\sc ii}~H line and 1700\,\AA\ (SDO/AIA). Then
we slowly scanned the chosen region and corrected its position along the
spectrograph slit, using the live image from the camera and taking the maximum
depth of the He\,{\sc i} 10\,830\,\AA\ line into consideration.
Figure\,\ref{fig:fac-scan} presents a spatial scan example for time series
No.\,8. Finally, the He\,{\sc i} 10\,830\,\AA\ intensity space--time diagrams
can be used to control the position of the observed object in the spectrograph
slit throughout a time series (Figure~\ref{fig:fac-data-overview}\,(b)).

\begin{table}
\caption[]{Facula data sets, lags between five-minute waves observed in two spectral lines}
\begin{tabular}{ccccccc}
\hline  \\
{\bf No.}&                           	
{\bf Date}&				
{\bf Disc}&				
{\bf T,\tabnote{total duration}}&	
{\bf t,\tabnote{sampling time (cadence)}}&	
{\bf Lag,}&	
{\bf Lines}				
\\
{\bf}&					
{\bf}&					
{\bf location}&				
{\bf min}&				
{\bf sec}&				
{\bf sec}&				
{\bf }					
\\ \hline
\\ 1 & 17 Jul 2010 &  19$^\circ$N  28$^\circ$W  & 150 & 3  &  78\,--\,100 & Si\,{\sc i}\,--\,He\,{\sc i}
\\ 2 & 04 Aug 2010 &  15$^\circ$N  05$^\circ$W  & 81  & 4  & -12\,--\,0   & Si\,{\sc i}\,--\,He\,{\sc i}
\\ 3 & 05 Aug 2010 &  20$^\circ$N  14$^\circ$W  & 96  & 3.5&  20\,--\,40  & Si\,{\sc i}\,--\,He\,{\sc i}
\\ 4 & 09 Aug 2010 &  17$^\circ$N  23$^\circ$E  & 136 & 3  &  40\,--\,87  & Si\,{\sc i}\,--\,He\,{\sc i}
\\ 5 & 09 Aug 2010 &  18$^\circ$N  22$^\circ$E  & 98  & 4  &  10\,--\,50  & Si\,{\sc i}\,--\,He\,{\sc i}
\\ 6 & 09 Aug 2010 &  15$^\circ$N  12$^\circ$E  & 68  & 3  &  68\,--\,77  & Si\,{\sc i}\,--\,He\,{\sc i}
\\ 7 & 14 Aug 2010 &  13$^\circ$N  05$^\circ$W  & 198 & 3  &  30\,--\,90  & Si\,{\sc i}\,--\,He\,{\sc i}
\\ 8 & 15 Aug 2010 &  24$^\circ$N  18$^\circ$E  & 102 & 3  &  33\,--\,39  & Si\,{\sc i}\,--\,He\,{\sc i}
\\ 9 & 16 Aug 2010 &  32$^\circ$N  00$^\circ$E  & 55  & 3  &  68\,--\,78  & Si\,{\sc i}\,--\,He\,{\sc i}
\\10 & 01 Jul 2003 &  04$^\circ$N  33$^\circ$E  & 42  & 10 &  -20\,--\,10 & Fe\,{\sc i}\,--\,H$\alpha$
\\11 & 18 Aug 2004 &  09$^\circ$N  17$^\circ$E  & 42  & 1  &  -8\,--\,20  & Fe\,{\sc i}\,--\,H$\alpha$
\\12 & 06 Jul 2010 &  19$^\circ$N  07$^\circ$W  & 92  & 2.5&  34\,--\,71  & Fe\,{\sc i}\,--\,H$\alpha$
\\13 & 17 Aug 2005 &  20$^\circ$S  30$^\circ$E  & 93  & 2  &   4\,--\,10  & Fe\,{\sc i}\,--\,Ba\,{\sc ii}
\\14 & 17 Aug 2005 &  20$^\circ$S  26$^\circ$E  & 88  & 2  &  -6\,--\,48  & Si\,{\sc i}\,--\,Ca\,{\sc ii}
\\
\\ \hline
\end{tabular}
\thetabnotes[]
\label{tbl:faculae}
\end{table}

\begin{table}
\caption[]{Sunspot data sets, lags between three-minute waves observed in two spectral lines}

\begin{tabular}{cccccccc}
\hline  \\
{\bf No.}&				
{\bf NOAA}&				
{\bf Date}&				
{\bf Disc}&				
{\bf T,\tabnote{total duration}}&	
{\bf t,\tabnote{sampling time (cadence)}}&	
{\bf Lag,}&				
{\bf Lines}				
\\
{\bf}&					
{\bf}&					
{\bf}&					
{\bf location}&				
{\bf min}&				
{\bf sec}&				
{\bf sec}				
\\ \hline
\\ 15 & 11251 & 18 Jul 2011 &  16$^\circ$N  10$^\circ$W  &  64 & 0.8  &  64\,--\,106  & Si\,{\sc i}\,--\,He\,{\sc i} 
\\ 16 & 11251 & 18 Jul 2011 &  16$^\circ$N  10$^\circ$W  &  52 & 0.8  &  20\,--\,64   & Si\,{\sc i}\,--\,He\,{\sc i} 
\\ 17 & 11251 & 19 Jul 2011 &  16$^\circ$N  20$^\circ$W  &  87 & 0.8  & -20\,--\,110  & Si\,{\sc i}\,--\,He\,{\sc i} 
\\ 18 & 11180 &  3 Apr 2011 &  24$^\circ$N  37$^\circ$W  &  67 & 5.5  &      35       & Si\,{\sc i}\,--\,He\,{\sc i} 
\\ 19 & 10613 & 20 May 2004 &  09$^\circ$S  03$^\circ$W  &  60 & 5    &      ---      & Fe\,{\sc i}\,--\,H$\alpha$ 
\\ 20 & 10661 & 18 Aug 2004 &  07$^\circ$N  19$^\circ$E  &  42 & 1    &      ---      & Fe\,{\sc i}\,--\,H$\alpha$ 
\\ 21 & 10791 & 27 Jul 2005 &  13$^\circ$N  02$^\circ$E  & 100 & 2    &      ---      & Fe\,{\sc i}\,--\,H$\alpha$ 
\\ 22 & 10791 & 27 Jul 2005 &  13$^\circ$N  02$^\circ$E  & 100 & 2    &      ---      & Fe\,{\sc i}\,--\,H$\alpha$ 
\\ 23 & 10794 &  5 Aug 2005 &  11$^\circ$S  16$^\circ$E  &  66 & 1    &      ---      & Fe\,{\sc i}\,--\,H$\alpha$ 
\\ 24 & 10810 & 21 Sep 2005 &  11$^\circ$N  30$^\circ$E  & 108 & 2    &      ---      & Fe\,{\sc i}\,--\,H$\alpha$ 
\\ 25 & 10810 & 23 Sep 2005 &  08$^\circ$N   0$^\circ$W  &  73 & 10   &      ---      & Fe\,{\sc i}\,--\,H$\alpha$ 
\\ 26 & 10657 & 13 Aug 2004 &  12$^\circ$N   5$^\circ$W  &  58 & 5    & -10\,--\,20   & Fe\,{\sc i}\,--\,Ba {\sc ii} 
\\ 27 & 10661 & 17 Aug 2004 &  09$^\circ$N  24$^\circ$E  &  42 & 0.5  &   0\,--\,22   & Fe\,{\sc i}\,--\,Ba {\sc ii} 
\\ 28 & 10661 & 18 Aug 2004 &  09$^\circ$N  13$^\circ$E  &  41 & 1    & -10\,--\,13   & Fe\,{\sc i}\,--\,Ba {\sc ii} 
\\ 29 & 10886 & 26 May 2006 &  08$^\circ$N  10$^\circ$E  &  67 & 2    & -12\,--\,54   & Si\,{\sc i}\,--\,Ca {\sc ii} 
\\ 30 & 10905 & 28 Aug 2006 &  06$^\circ$S  10$^\circ$W  &  27 & 0.5  &      ---      & Si\,{\sc i}\,--\,Ca {\sc ii} 
\\ 31 & 10908 & 11 Sep 2006 &  14$^\circ$S  09$^\circ$E  &  25 & 0.5  &      ---      & Si\,{\sc i}\,--\,Ca {\sc ii} 
\\ \\ \\ \hline
\end{tabular}
\thetabnotes[]
\label{tbl:spots}
\end{table}

\begin{table}
\caption[]{Spectral line formation heights}
\begin{tabular}{cccc}
\hline  \\
{\bf Line }&                           	
{\bf Height, }&				
{\bf Atmosphere model}&				
{\bf Reference}	
\\
{\bf}&					%
{\bf Mm }&				%
{\bf}&				
{\bf }					%
\\ \hline
\\ He\,{\sc i} 10\,830\,\AA &  $\sim{2}$ & FALC & Avrett \etal, \citeyear{avrett1994-helium}
\\ Si\,{\sc i} 10\,827\,\AA &  0.3   &  model-M & \cite{bard2008si-he}
\\ Si\,{\sc i} 10\,827\,\AA &  0.5  & FALC & \cite{bard2008si-he}
\\ Ba\,{\sc ii} 4554\,\AA\ &  0.64  &  3D MHD  & Shchukina \etal, \citeyear{shchukina2009}
\\ Fe\,{\sc i}  4551\,\AA\ & 0.14  & HOLMU & \cite{gurtovenko1989}
\\ Ca\,{\sc ii} 8542\,\AA\ &  1.2-1.5 &  HSRA  & \cite{mein1980}
\\ Ca\,{\sc ii} 8542\,\AA\ &  1.0-1.5 &  3D MHD & \cite{leenaarts2009-ca}
\\ H$\alpha$ 6563\,\AA\  & 1.5-2  & VAL  &  Vernazza \etal, \citeyear{vernazza81}
\\ H$\alpha$ 6563\,\AA\ & 1.0-1.6  & 3D MHD  &  Carlsson \etal,  \citeyear{leenaarts2012-halpha}
\\ Fe\,{\sc i}  6569\,\AA\ & 0.15  & Bilderberg & \cite{parnell69}

\\ \hline

\end{tabular}
\label{tbl:lineform}
\end{table}

\begin{figure} 
\centerline{
\includegraphics[width=5.5cm]{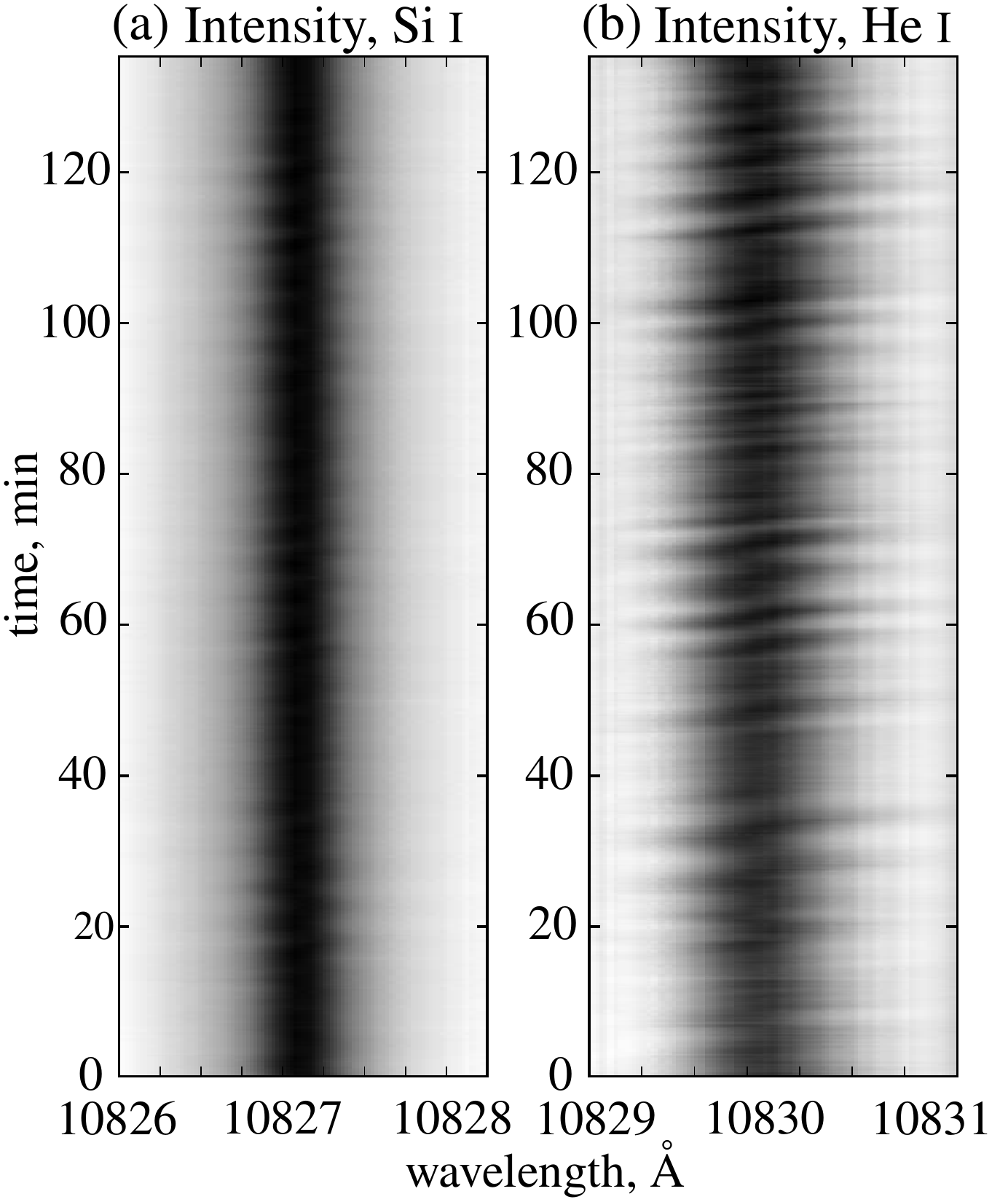}
}
\caption{Time series No.\,4. Temporal intensity evolution for
the (a) Si\,{\sc i} 10\,827\,\AA\ and (b) He\,{\sc i} 10\,830\,\AA\ lines.
        }
\label{fig:fac-stokes}
\end{figure}

\section{Measurements of the Time Delays Between LOS-velocity oscillations
at Different Heights of the Solar Atmosphere}
\subsection{Faculae}

\par Figure \ref{fig:fac-stokes} demonstrates the quality of the data.
Table~\ref{tbl:faculae} presents the details about the time series and the
measurement results for the time lag of the chromospheric LOS velocity relative
to the photospheric velocity. The spatial coherency of the LOS-velocity
oscillations in our measurements is at least about 2$''$. The five-minute band
oscillation power spectra often present a fine structure. At first, we
determined frequency peaks that dominate both in the photosphere and in the
chromosphere. Then the LOS-velocity signals, averaged over 2$''$, were subjected
to frequency filtration in the narrow band, usually 1~mHz, centred at the
determined frequency peak. The mean amplitude of the whole photospheric signal
series was equalised to match the chromospheric signal. The lag between the
signals was determined using cross-correlation and was confirmed visually. As
can be seen from Table~\ref{tbl:faculae}, the LOS-velocity five-minute
oscillation chromospheric signal lags behind the photospheric signal for most of
the faculae. The time lag generally does not exceed 100\,s. A negative lag
(downward-propagating wave) was detected for some of the time series ({\it e.g.}
No. 2). Waves propagating downward were reported by \cite{finsterle2004},
\cite{mcateer2003}. The corresponding wave packets at two levels can
differ in detail, and different wave trains can reveal different lags.

\begin{figure}
\includegraphics[width=\textwidth]{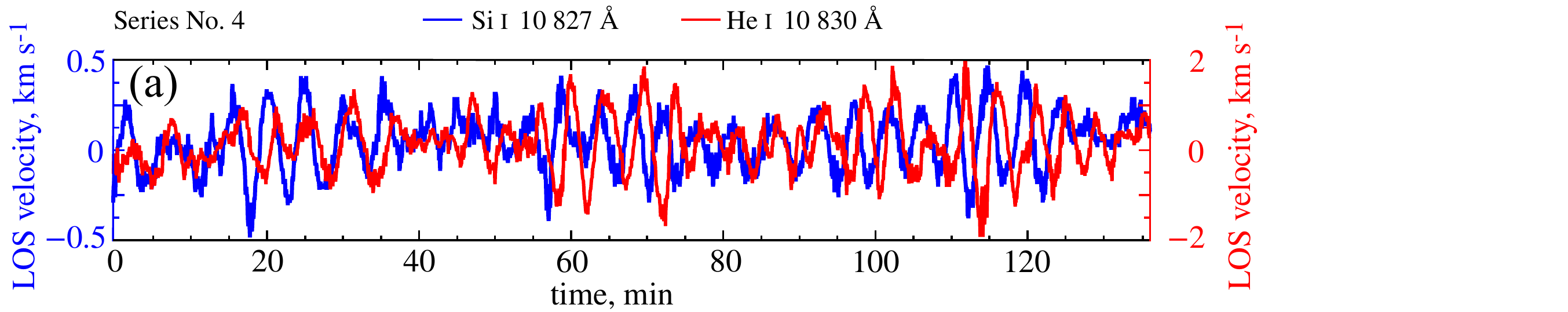}
\caption{Time series No. 4. Original (unfiltered) LOS-velocity signals: facular
photosphere (Si\,{\sc i} 10\,827\,\AA, blue line) and chromosphere
(He\,{\sc i} 10\,830\,{\AA}, red line).
        }
\label{fig:fac-he-si-raw}
\end{figure}

\begin{figure}
\includegraphics[width=\textwidth]{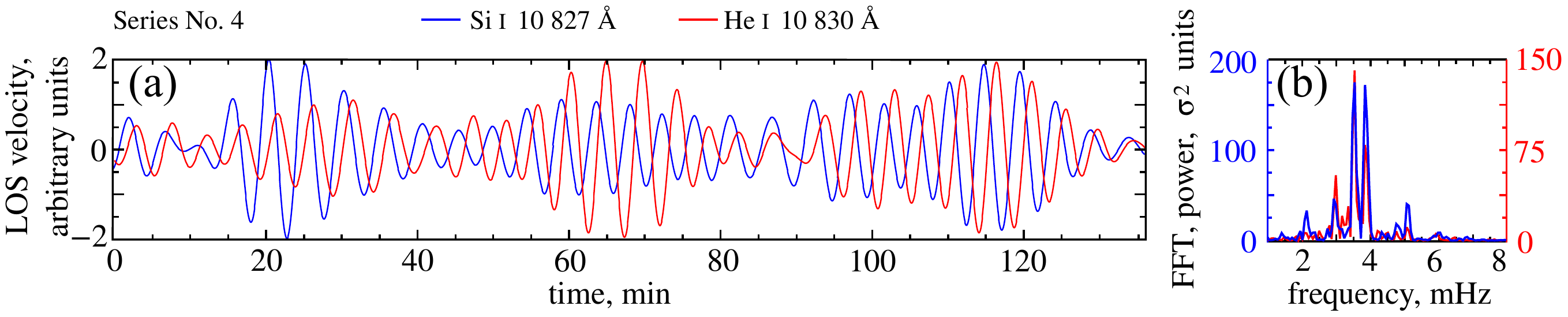}
\caption{Time series No. 4. (a) Filtered LOS-velocity signals: facular
photosphere (Si\,{\sc i} 10\,827\,\AA, blue line) and chromosphere
(He\,{\sc i} 10\,830\,{\AA}, red line). 3\,--\,4 mHz filtration band.
(b) Power spectrum for the LOS-velocity signals at two height levels.
        }
\label{fig:fac-he-si}
\end{figure}

\begin{figure}
\centerline{
\includegraphics[width=5cm]{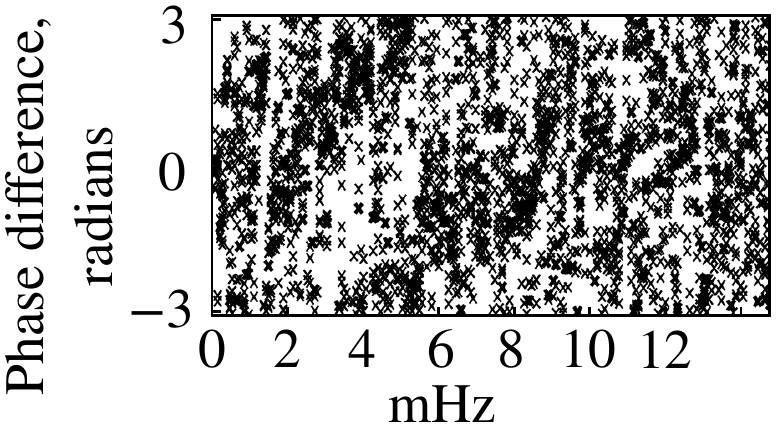}
}
\caption{Time series No.\,4.
The FFT phase difference of LOS-velocity oscillations
($V_{Si\,{I}} - V_{He\,{I}}$) for each facula element along the entrance
slit and for a specific frequency.
        }
\label{fig:fac-phase}
\end{figure}

\par Figure~\ref{fig:fac-he-si-raw} presents the LOS velocity variations for a
single position along the slit. The forms of the photospheric and chromospheric
signals are very similar. Five-minute oscillations forming three wave trains are
visible at both levels. These signals were filtered to determine the time lag
more reliably (Figure~\ref{fig:fac-he-si}). The frequency filter pass band was
set to be 1 mHz, centred at 3.5\,mHz. Narrowing the pass band did not
significantly affect the structure of the profiles. After the
filtration the wave trains became more pronounced. At the same time, this
revealed noticeable differences in the oscillations within a single train
between the photosphere and the chromosphere. The cross-correlation between the
photospheric and chromospheric signals is highest (0.97) for an 87\,s lag. This
value corresponds to the single spatial domain of 2$''$ along the slit and is
the longest phase lag detected for this time series. A lowest lag value of 40\,s
was measured for another spatial point along the slit. Table~\ref{tbl:faculae}
presents these values as 40\,--\,87\,s for series No. 4. Similarly, the lag
values are given for other time series. For the majority of the time series, the
LOS velocity power spectrum in the Si\,{\sc i} 10\,827\,\AA\ line is very
similar to that in the He\,{\sc i} 10\,830\,{\AA} line.

\par To diagnose the wave propagation, one can use a phase difference plot
\citep{lites1992a}. Figure\,\ref{fig:fac-phase} shows such a plot for time series
No.\,4. Each ``x'' marks the phase difference of the LOS-velocity oscillations
($V_{Si\,{I}} - V_{He\,{I}}$) for each facula element along the entrance slit
and particular frequency. Weak indications of propagating oscillations can be
seen at 3\,mHz, where one can see a jump in the phase difference. The time delay
between the oscillations of these frequencies can be roughly estimated as 66\,s.
This value is within the range between the lowest and highest values of the
phase lags detected by direct comparison of the filtered LOS-velocity signals
(Table~\ref{tbl:faculae}, series No.\,4).

\par Line-of-sight velocity signals were obtained from different spectral lines
formed at different heights within the solar atmosphere. The phase speed can be
determined for an upward propagating wave. For Si\,{\sc i} 10\,827\,\AA\ and
He\,{\sc i} 10\,830\,\AA\ the height difference is 1500 km
\citep{avrett1994-helium,bard2008si-he,centeno09}, see Table\,\ref{tbl:lineform}.
If we observe the waves that travel upward at the photospheric sound speed
(4\,--\,6 {$\mathrm {km\,s^{-1}}$}), then the lag is to be about 300\,s. There
are no such lag values in our measurements (see Table~\ref{tbl:faculae}). The
50\,--\,100\,s lags correspond to a speed of 15\,--\,30 {$\mathrm
{km\,s^{-1}}$}. This is much higher than the speed of sound.

\begin{figure}
\includegraphics[width=\textwidth]{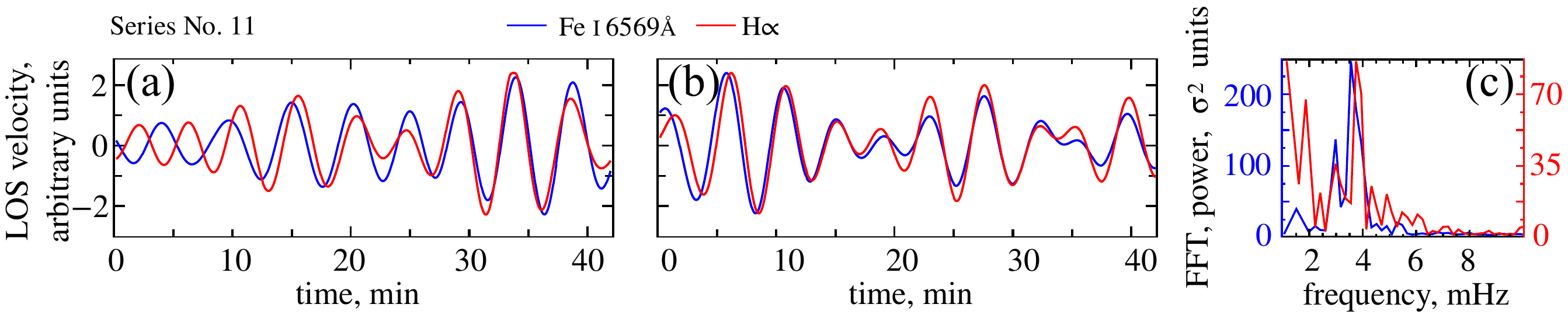}
\caption{Time series No.\,11. Filtered LOS-velocity signals for
two positions on the slit (a) and (b), 3\,--\,4 mHz filtration band.
(c) Power spectra for the original signals shown in (a).
Photospheric signal Fe\,{\sc i} 6569\,\AA\ -- blue line;
chromospheric signal H$\alpha$ 6563\,\AA\ -- red line.
        }
\label{fig:fac-fe-ha}
\end{figure}

\par In contrast to the Si\,{\sc i} 10\,827\,\AA\ and He\,{\sc i} 10\,830\,\AA\
pair, power spectra for facular oscillations derived from the Fe\,{\sc i}
6569\,\AA\ and H$\alpha$ 6563\,\AA\ lines differ significantly in the
low-frequency (1\,--\,2~mHz) and high-frequency (5\,--\,7~mHz) regions. The
power of chromospheric oscillations at 1\,--\,2 mHz frequencies obtained in
H$\alpha$ is often comparable to or exceeds the power of five-minute oscillations. A
similarity between the spectra is observed only in the central part of the range
(2.5\,--\,4.5~mHz). The oscillation power in this spectral range is
considerably higher than the noise level (Figure~\ref{fig:fac-fe-ha}~(c)). The
highest cross-correlation for the signals in Figure~\ref{fig:fac-fe-ha}~(a)
corresponds to an 8\,s negative lag. On the other hand, the 10\,--\,20 minute
range clearly shows a positive delay. The signals for another slit position of
the same time series are shown in Figure~\ref{fig:fac-fe-ha}~(b). The delay for the
first wave train is 20\,s. The second wave train shows no delay between the
photosphere and the chromosphere. The other time series obtained in these lines
show a similar behaviour of the photospheric and chromospheric signals. The
Fe\,{\sc i} 6569\,\AA\ line formation depth is 150\,--\,200 km \citep{parnell69}
and that of H$\alpha$ 6563\,\AA\ is 1550\,--\,2000
\citep{vernazza81,leenaarts2012-halpha}. The longest 71\,s lag detected
corresponds to a phase speed of about 21~{$\mathrm {km\,s^{-1}}$}.

\begin{figure}
\includegraphics[width=\textwidth]{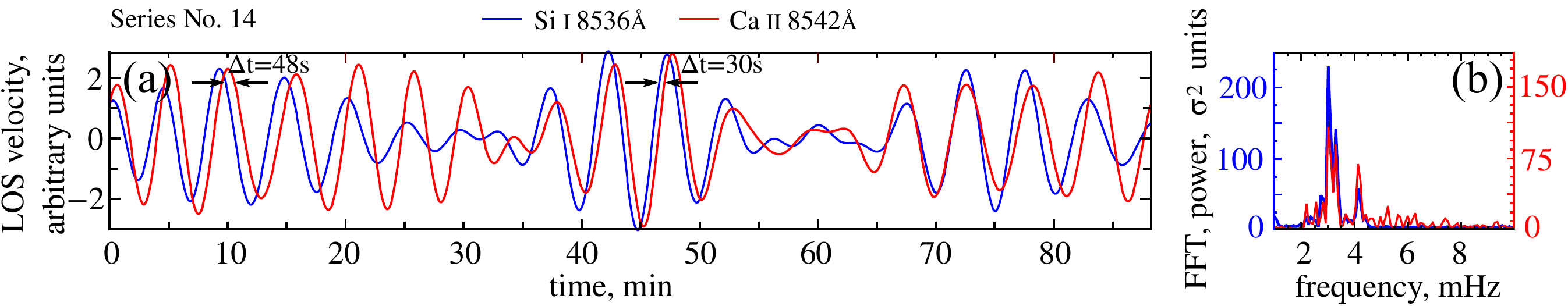}
\caption{Time series No. 14. (a) Photospheric and chromospheric
LOS-velocity signals filtered in the 2.5\,--\,3.5 mHz band. (b) Corresponding
power spectra of the LOS-velocity signals for two atmospheric levels.
Photosphere Si\,{\sc i} 8536\,\AA\ -- blue line;
chromosphere Ca\,{\sc ii} 8542\,\AA\ -- red line.
        }
\label{fig:fac-si-ca}
\end{figure}
\begin{figure}
\includegraphics[width=\textwidth]{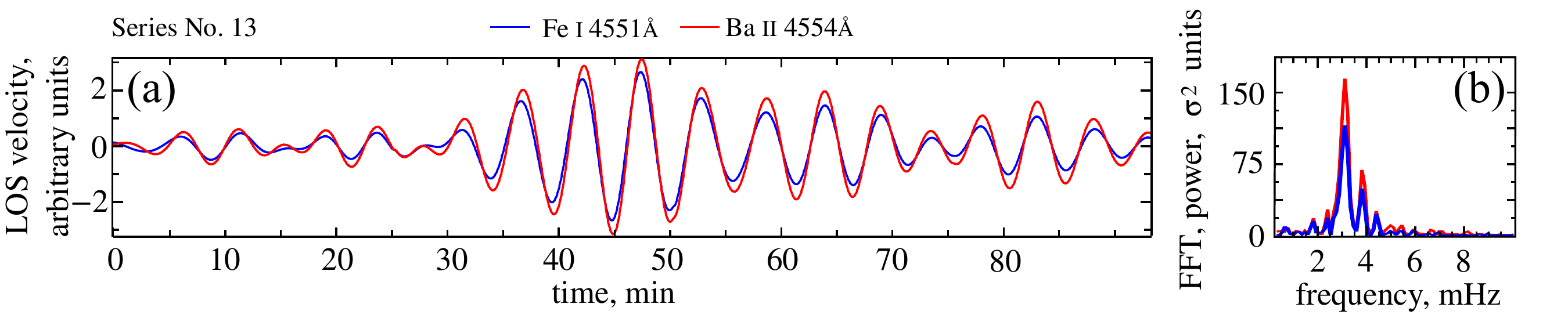}
\caption{Time series No. 13: (a) LOS-velocity signals filtered in the
2.5\,--\,3.5 mHz band. (b) Corresponding power spectra of the LOS-velocity
signals for two atmospheric levels.
Photosphere Fe\,{\sc i} 4551\,\AA\ -- blue line;
temperature minimum Ba\,{\sc ii} 4554\,\AA\ -- red line.
        }
\label{fig:fac-ba-fe}
\end{figure}
\par The character of the oscillations detected in Si\,{\sc i} 8536\,\AA\ and
Ca\,{\sc ii} 8542\,\AA\ is similar to that observed in Fe\,{\sc i} 6569\,\AA\
and H$\alpha$ 6563\,\AA. The first wave train in the photosphere differs notably
from that in the chromosphere (Figure~\ref{fig:fac-si-ca}~(a)). If we assume
that we observe the same wave train at both levels, the estimated phase delay is
48\,s. The second wave train is clearly identified at both levels and the
chromospheric signal lags behind the photospheric signal by 30\,s. The third
wave train shows an ambiguous delay that gradually changes from a slightly
negative to a positive one. At the same time, oscillations in the five-minute band
dominate at both levels, and their power is significantly higher than the noise
level (Figure~\ref{fig:fac-si-ca}~(b)). The estimated formation depth
of the Ca\,{\sc ii} 8542\,\AA\ line is within the 1200\,--\,1500\,km range
\citep{mein1980,leenaarts2009-ca}. With a 1000\,km height difference, this
roughly corresponds to a lowest phase speed of 20\,--\,30 {$\mathrm
{km\,s^{-1}}$}, which is much higher than the speed of sound in the photosphere.

\par In this context it is interesting to analyse the situation for the solar
atmospheric levels that are close in height. For example, Fe\,{\sc i} 4551\,\AA\
and Ba\,{\sc ii} 4554\,\AA\ provide us with such an opportunity. The core of the
Ba\,{\sc ii} 4554\,\AA\ line forms at 640~km \citep{shchukina2009}. The depth of
the Fe\,{\sc i} 4551\,\AA\ line formation is 140~km \citep{gurtovenko1989}.
Hence, the maximal height difference is 500~km. As can be seen from
Figure~\ref{fig:fac-ba-fe}, the signals for both levels are almost identical,
except for an 8\,s time lag and the amplitude difference by a factor of 2. The
whole facula (average over the slit) demonstrates the same behaviour. The
longest 10\,s lag (see Table\,\ref{tbl:faculae}) corresponds to a phase speed of
50 {$\mathrm {km\,s^{-1}}$}.

\subsection{Sunspots}
\par Waves in sunspots have been a matter of debate since the 1970s. In
contrast to the faculae, three-minute oscillations dominate in the umbra sunspot
chromosphere. The measured phase delays are still difficult to interpret ({\it
e.g.} \cite{lites1992a}). In the observations reported by different authors
three-minute oscillations manifest themselves as standing waves or as
upward-propagating waves. \cite{lites1985a} found that three-minute oscillations
of the LOS velocity in the photosphere have the character of vertical standing waves.
The early research \citep{uexkuell1983a} estimated the wave's upward-propagating
speed to be 10\,--\,25 {$\mathrm {km\,s^{-1}}$}, depending on the frequency.
\cite{giovanelli1978a} measured the lag between photospheric and
chromospheric oscillations to be about 26\,s (Fe\,{\sc i}
5233\,\AA\,--\,H$\alpha$ pair). \cite{kentischer1995} reported a 40\,s
time lag between the intensity signals of the H$\alpha$ and Ca\,{\sc ii}\,K
lines. This lag corresponds to a vertical phase speed of 7 {$\mathrm
{km\,s^{-1}}$}. \cite{kobanov2011} determined a non-constant phase delay,
changing from 20 to 140\,s (Fe\,{\sc i} 6569\,\AA\,--\,H$\alpha$ lines). A
comparison of the optical and radio observations showed a time delay of 50\,s of
the radio signal with respect to the optical signal \citep{abramov2011}. Long
$\sim$300\,s phase delays (low propagation speed) for the Si\,{\sc i}
10\,827\,{\AA} and He\,{\sc i} 10\,830\,{\AA} pair were reported by
\cite{centeno09}.

\begin{figure} 
\centerline{
\includegraphics[width=11.5cm]{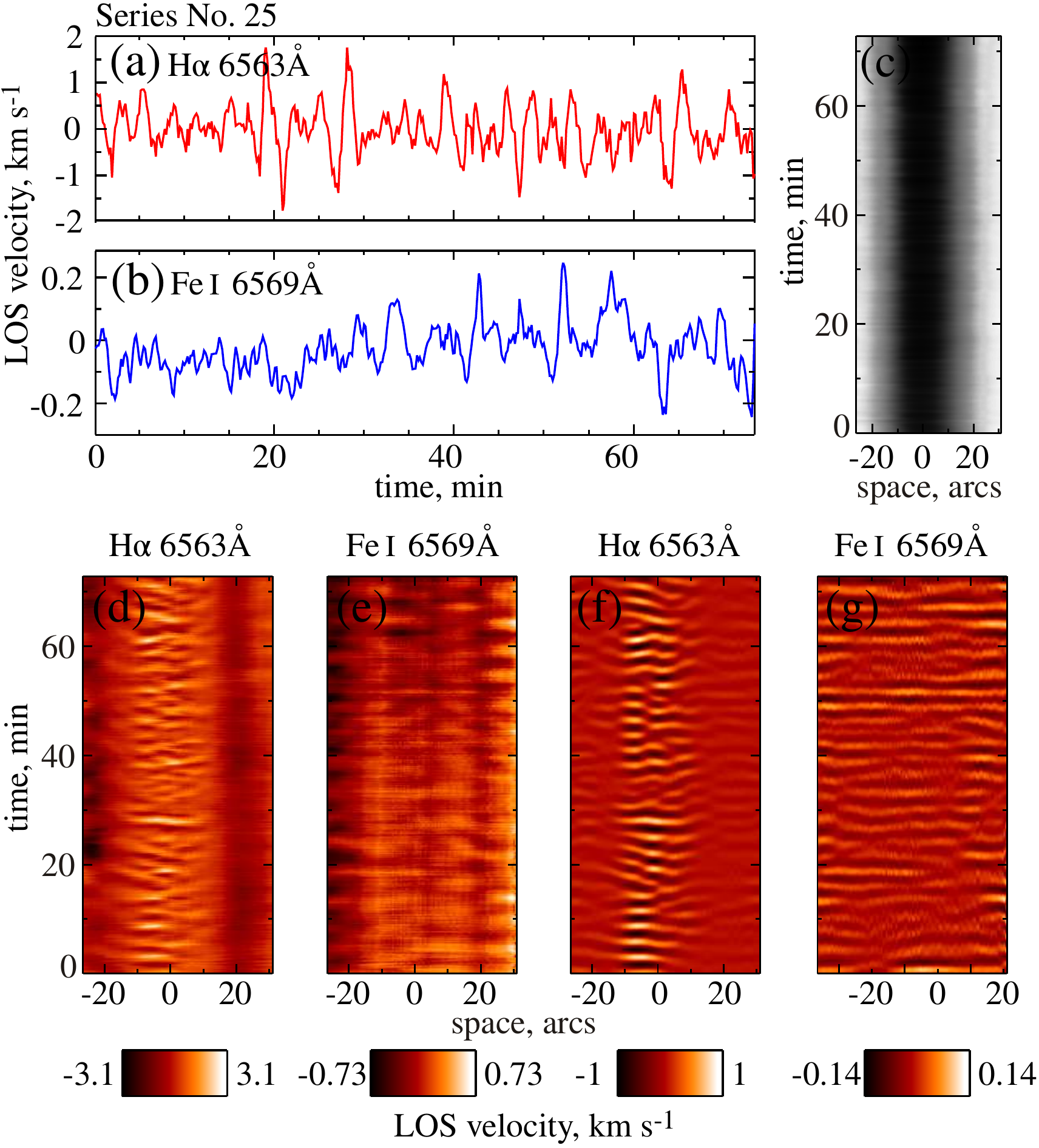}
}
\caption{Time series No.\,25. Overview of the data. Panels (a) and (b):
unfiltered LOS-velocity signals for umbra centre slit position. Panel (c):
intensity signal in continuum; (d) and (e): unfiltered diagrams of the LOS-velocity
signals for the whole slit; (f) and (g): corresponding diagrams after the wavelet
filtration (4.6\,--\,7.2\,mHz band).
        }
\label{fig:spot-data-overview}
\end{figure}
\begin{figure} 
\centerline{
\includegraphics[width=8cm]{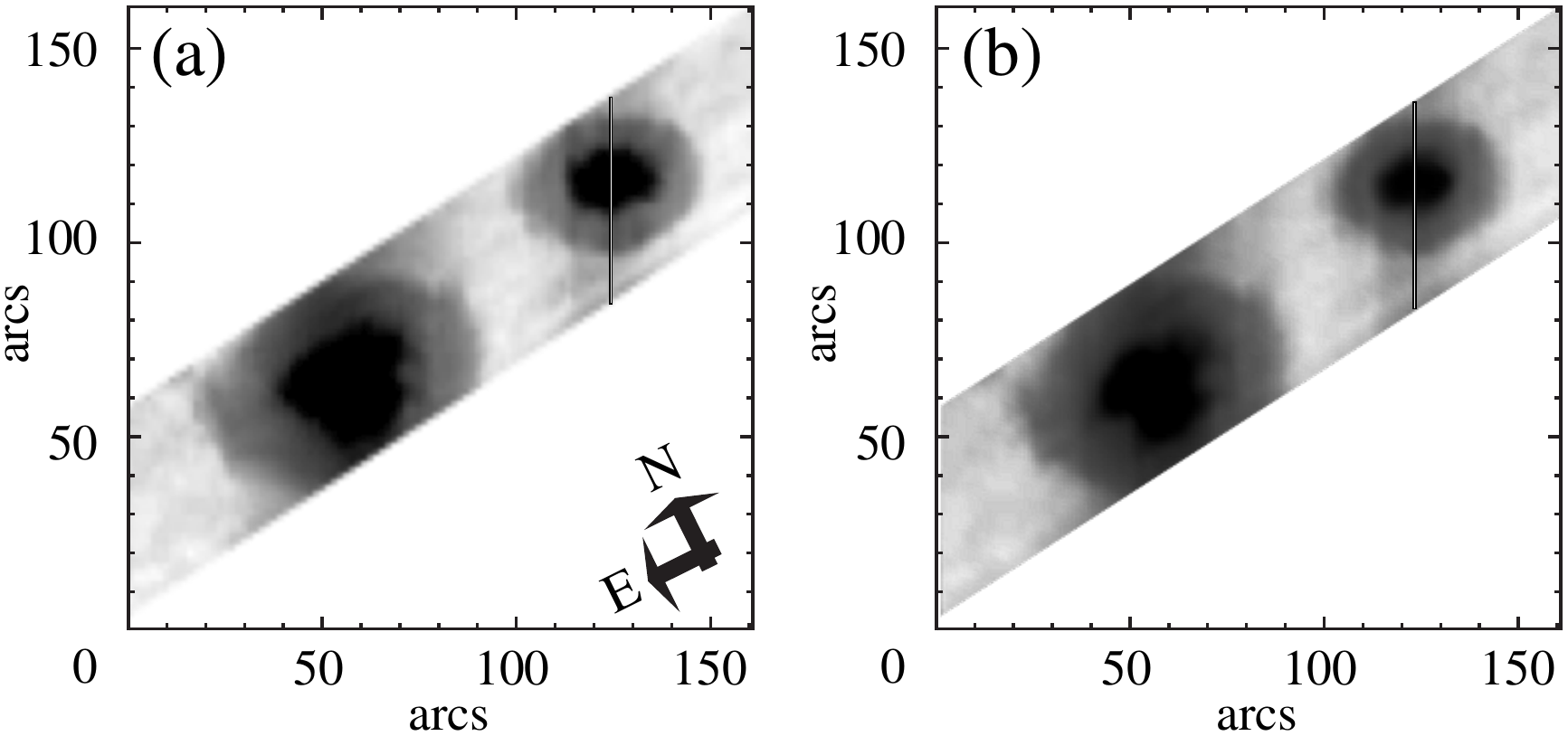}
}
\caption{Overview scan of the active region that is observed in time series
No.\,20. Panel (a): before the time sequence and (b) after the time sequence.
        }
\label{fig:spot-scan}
\end{figure}

\begin{figure} 
\centerline{
\includegraphics[width=3.1cm]{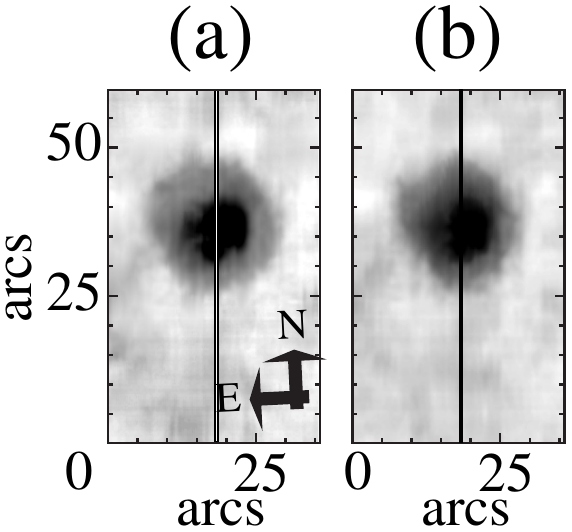}
}
\caption{Overview scan of the active region that is observed in time series
No.\,16. Panel (a): before the time sequence and (b) after the time sequence.
        }
\label{fig:spot-scan-s15}
\end{figure}

\begin{figure} 
\centerline{
\includegraphics[width=11.5cm]{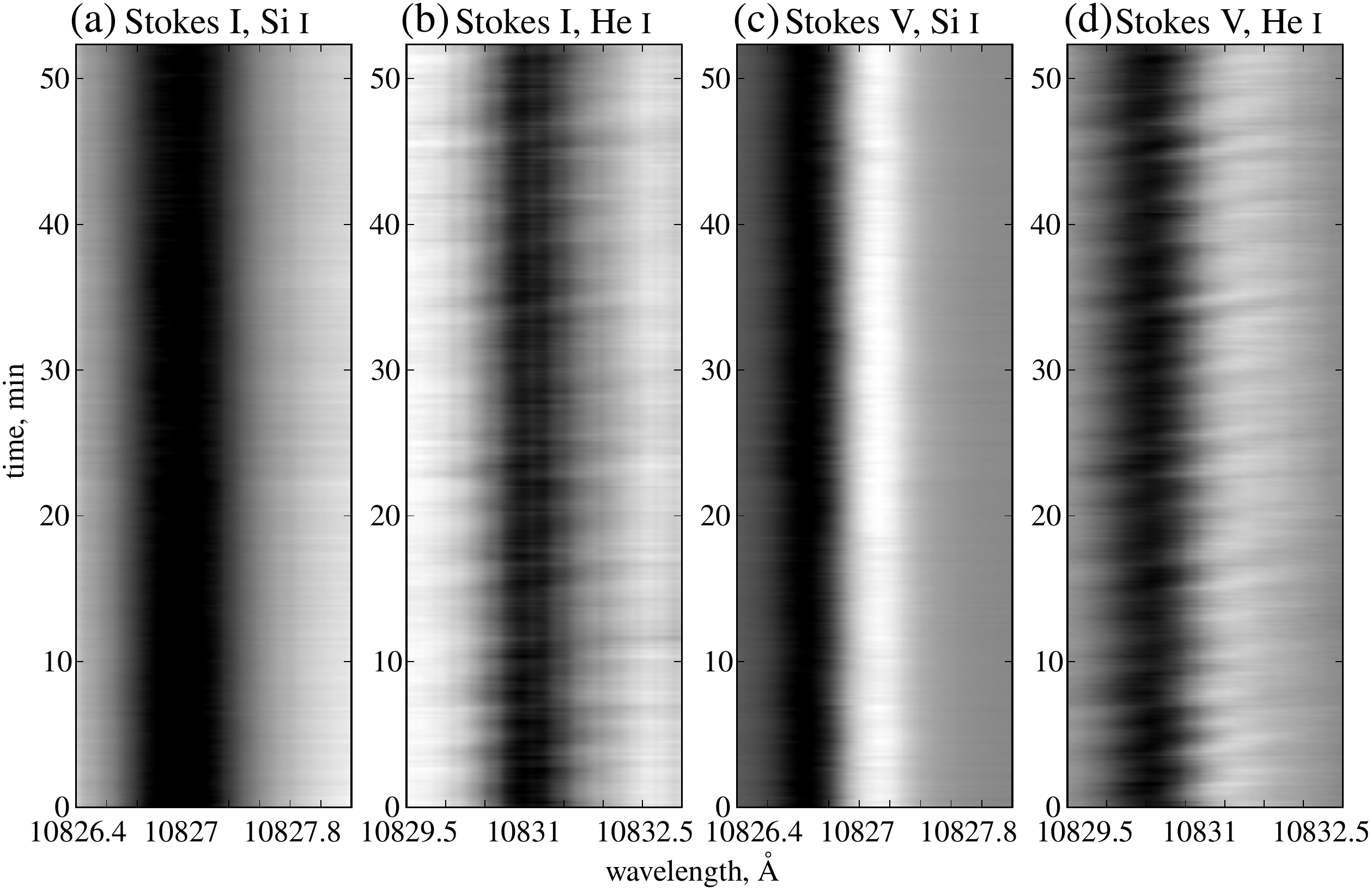}
}
\caption{Time series No.\,16. Temporal evolution of Stokes I and Stokes V for
Si\,{\sc i} 10\,827\,\AA\ (a) and (c) and He\,{\sc i} 10\,830\,\AA\ (b) and (d).
        }
\label{fig:spot-stokes}
\end{figure}

\par Here we present delays measured for 17 time series (12 sunspots,
Table~\ref{tbl:spots}). As an example, Figure~\ref{fig:spot-data-overview}
presents the data for time series No.\,25. The sunspot position is stable
throughout the whole time series (Figure~\ref{fig:spot-data-overview}~(c)).
Panels (a) and (b) demonstrate unfiltered LOS-velocity signals for the umbra
centre. Unfiltered space--time diagrams of the LOS-velocity signals for the
whole slit can be found in panels (d) and (e). The corresponding diagrams after
wavelet filtration (4.6\,--\,7.2\,mHz band) are also provided (panels (f) and (g)).
When possible, the time series were accompanied by the overview scan of the observed
spot. Figure~\ref{fig:spot-scan} shows an example of such a scan
for time series No.\,20 (continuum intensity). The spectrograph slit position
is marked by the vertical line. In some time series a slit-jaw image taken through
an H$\alpha$ filter was used to simultaneously monitor the target
position. The overview scan along with the Stokes I and V diagrams for the
sunspot series No.\,16 analysed below are shown in
Figures~\ref{fig:spot-scan-s15} and ~\ref{fig:spot-stokes}, respectively.

\begin{figure}
\centerline{
\includegraphics[width=5cm]{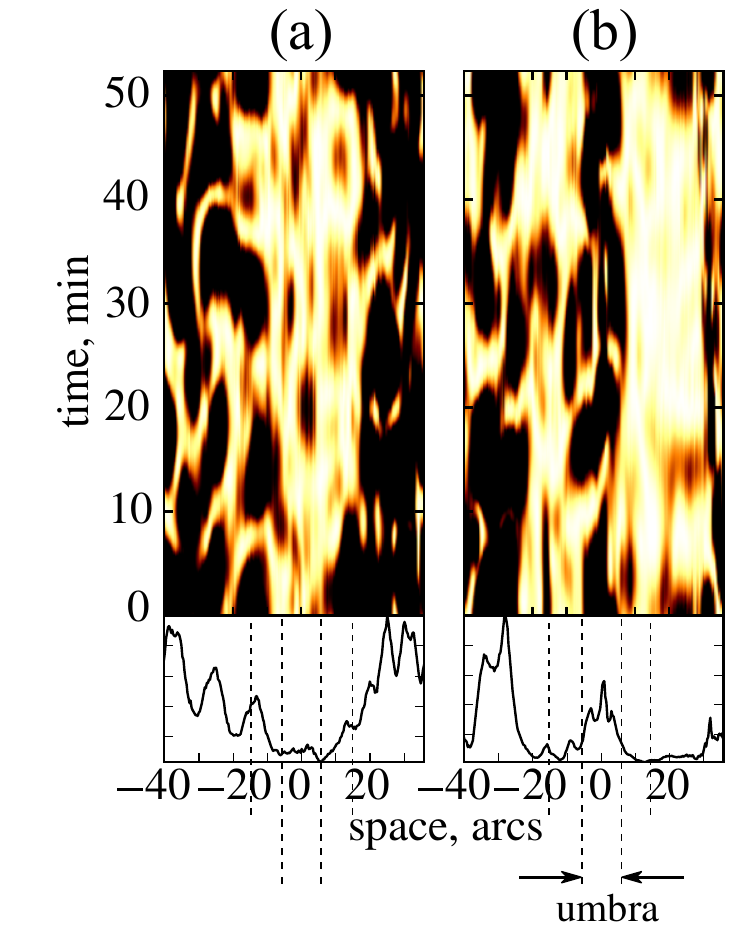}
}
\caption{Time series No.\,16. Power distribution of the three-minute
oscillations for the umbral centre (a)~photosphere and (b) chromosphere.
Dashed lines denote the penumbra boundaries.
        }
\label{fig:spot-2d-wlt-power}
\end{figure}

\begin{figure}
\centerline{
\includegraphics[width=4cm]{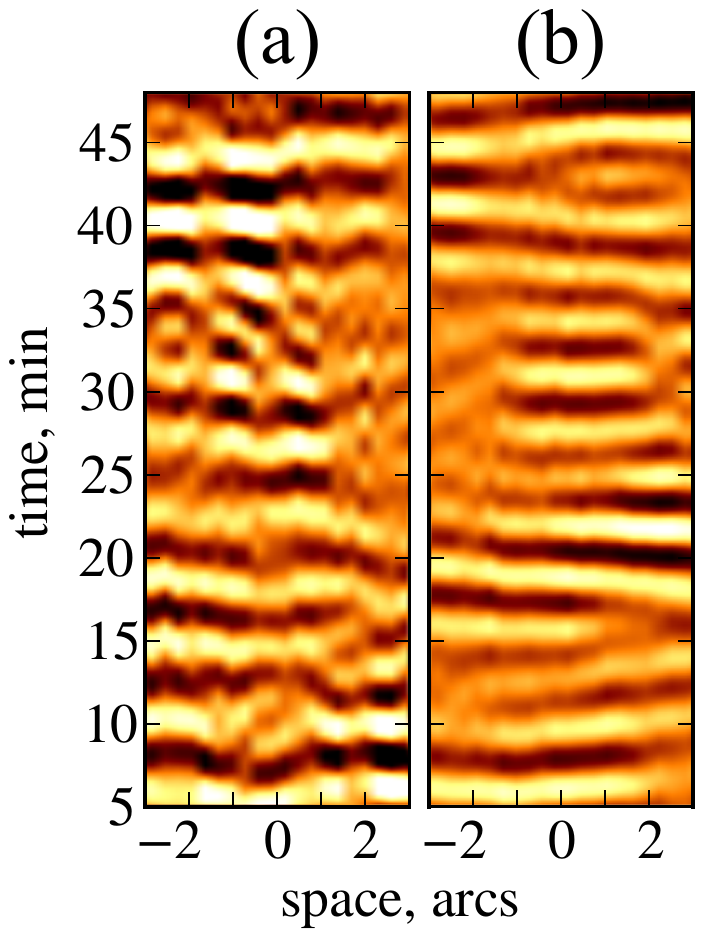}
}
\caption{Time series No.\,16. Space--time diagrams of LOS-velocity
signals (Stokes-V zero-crossing, 4.6\,--\,5.6\,mHz filtration) for
(a) the photosphere and (b) the chromosphere.
        }

\label{fig:spot-2d-si-he}
\end{figure}

\par Three-minute photospheric oscillations in the spot umbra are suppressed compared
to the neighbouring regions, whereas the chromospheric oscillations are enhanced.
This is valid not only for the Fe\,{\sc i} 6569\,\AA\,--\,H$\alpha$ lines
\citep{kobanov2011}, but also for the Si\,{\sc i} 10\,827\,\AA\,--\,He\,{\sc i}
10\,830\,{\AA} pair (see Figure~\ref{fig:spot-2d-wlt-power}). The distribution
of the three-minute oscillation power spectrum across the slit in
Figure~\ref{fig:spot-2d-wlt-power}\,(a) has the lowest value in the umbral
photosphere. The chromospheric signal is highest at the same slit position.
The filtration was performed for the 4.6\,--\,5.6\,mHz band. Series 19\,--\,25
obtained in the Fe\,{\sc i} 6569\,\AA\,--\,H$\alpha$ lines show the power of
the photospheric three-minute oscillations to be slightly higher than the noise level (exceeds
3$\sigma^2$ level). For most of these series it was not possible to find a good
correspondence between the three-minute oscillations in the photosphere and the
chromosphere. The cross-correlation between the signals is below 0.3. The
identification of wave trains at both levels was complicated. For these cases
the ``Lag'' column of Table\,\ref{tbl:spots} contains a dash for the corresponding
series.

\begin{figure}
\centerline{
\includegraphics[width=11.5cm]{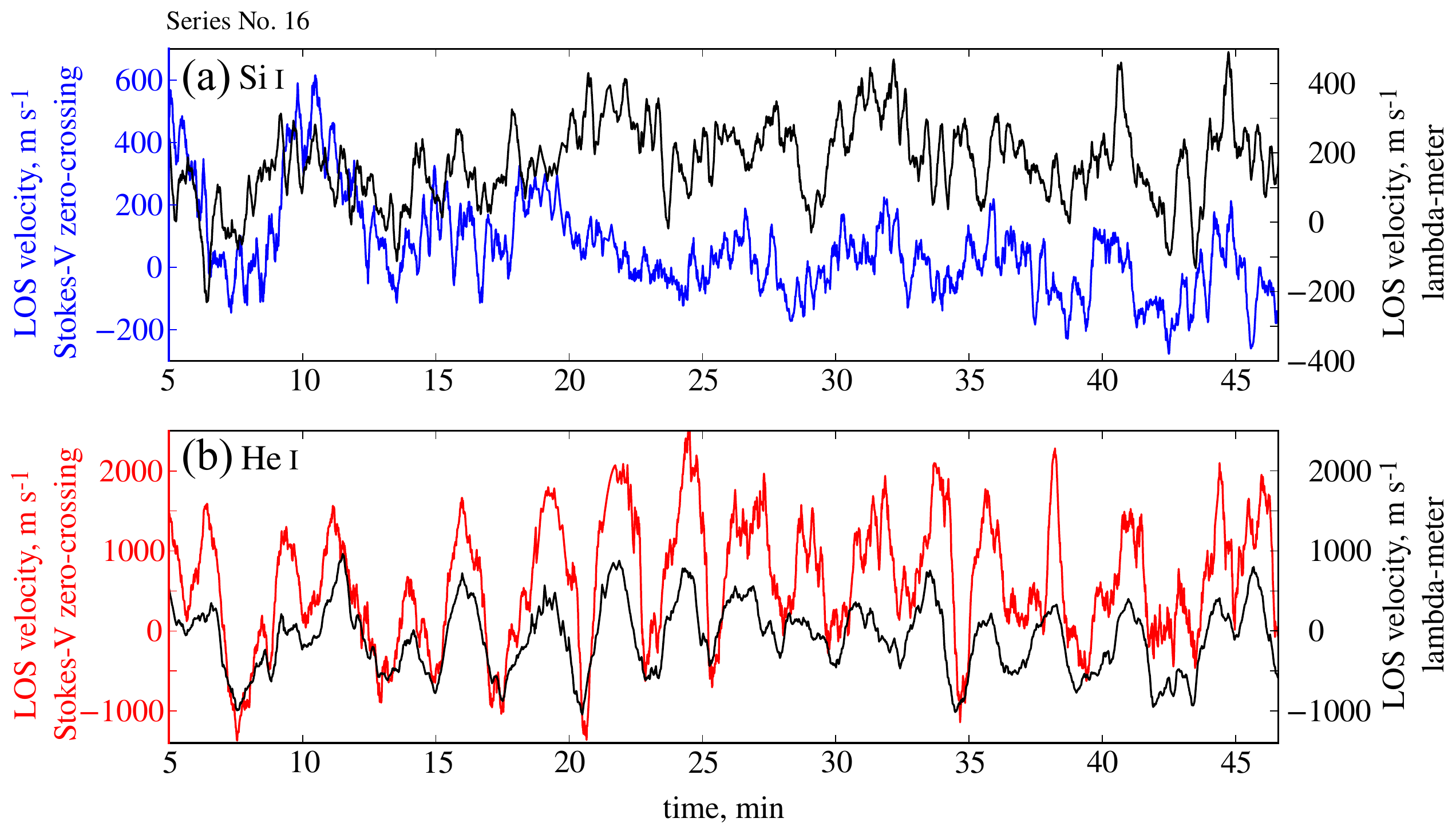}
}
\caption{
Time series No.\,16. The area corresponds to 1$''$ along the slit. The
unfiltered LOS velocities derived with the lambda-meter technique (black lines)
and with tracking Stokes-V zero-crossing (blue and red): (a) Si\,{\sc i}
10\,827\,\AA, (b) He\,{\sc i} 10\,830\,{\AA}.
}
\label{fig:spot-he-lme-vmp-raw-signal}
\end{figure}

\par
\begin{figure}
\centerline{
\includegraphics[width=11.5cm]{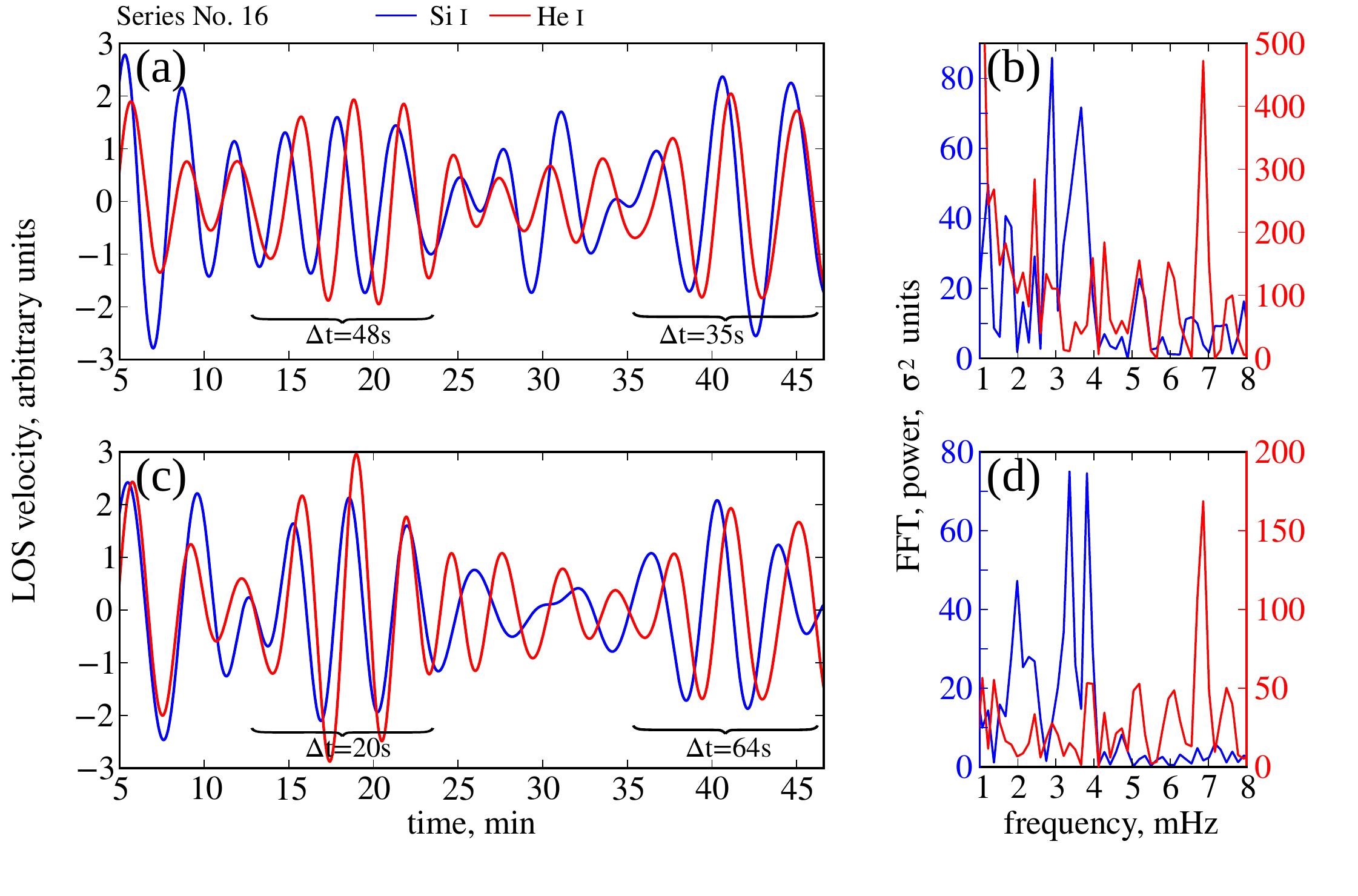}
}
\caption{
Time series No.\,16. LOS-velocity signals (4.5\,--\,5.6\,mHz filtration band)
derived with (a) the lambda-meter technique and (c) with Stokes-V zero-crossing. The
LOS-velocity power spectra for two atmospheric levels, derived witha (b) the
lambda-meter technique and (d) with Stokes-V zero-crossing. Photosphere -- Si\,{\sc
i} 10\,827\,\AA, blue line; chromosphere -- He\,{\sc i} 10\,830\,{\AA}, red
line. The position along the slit is the same as for
Figure\,\ref{fig:spot-he-lme-vmp-raw-signal}.
}
\label{fig:spot-si-he-lme-vmp}
\end{figure}

\begin{figure}
\centerline{
\includegraphics[width=10cm]{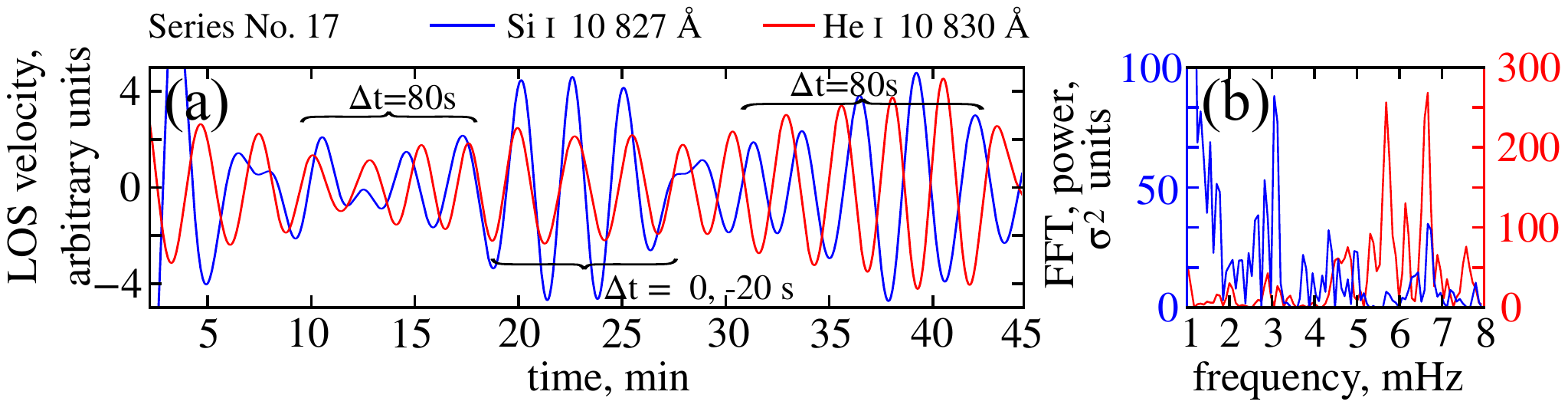}
}
\caption{Time series No.\,17, umbral region of 1$''$. (a) LOS velocity
(Stokes-V zero-crossing) filtered in the 6.0\,--\,7.0\,mHz band.
(b) The LOS-velocity power spectra for two atmospheric levels.
Photosphere -- Si\,{\sc i} 10\,827\,\AA, blue line;
chromosphere -- He\,{\sc i} 10\,830\,{\AA}, red line.
        }
\label{fig:spot-si-he-vmp-good}
\end{figure}

 The Si\,{\sc i} 10\,827\,\AA\ line is formed higher in the solar atmosphere
than that of Fe\,{\sc i} 6569\,\AA. The photospheric three-minute LOS-velocity
oscillation power of the former is considerably higher than the power detected
in the latter (see Figure~\ref{fig:spot-si-he-lme-vmp}\,(b) and (d))). In this
case it is possible to identify wave trains at both atmospheric levels for the
Si\,{\sc i} 10\,827\,\AA\,--\,He\,{\sc i} 10\,830\,{\AA} pair. LOS-velocity
signals, filtered in the 4.6\,--\,5.6\,mHz band, are shown in
Figure~\ref{fig:spot-2d-si-he}~(a) and (b). The highest cross-correlation of
0.65 between the two images can be achieved by a 51-s backward shift of the
He\,{\sc i} 10\,830\,{\AA} image. The unfiltered LOS-velocity signals exhibit
photospheric oscillations in the five-minute band
(Figure\,\ref{fig:spot-he-lme-vmp-raw-signal}~(a)) and the chromospheric oscillations in
the three-minute band (Figure\,\ref{fig:spot-he-lme-vmp-raw-signal}~(b)). The signals are
shown for the single position along the slit $-2''$ away from the umbra centre.
Figure~\ref{fig:spot-si-he-lme-vmp}~(a) and (c) presents these signals filtered
to reveal the three-minute oscillations. The upper panels represent the results derived
from the lambda-meter technique, the bottom panels depict the signals calculated by tracking
Stokes-V zero-crossing. The right panels show the corresponding power spectra of
the original (unfiltered) signals. The wavelet filtration was performed within
the 4.6\,--\,5.6\,mHz range, where the three-minute oscillation power is higher than
the noise level for all four signals (Figure~\ref{fig:spot-si-he-lme-vmp}~(b)
and (d)). One can see two wave trains in Figure~\ref{fig:spot-si-he-lme-vmp} (a)
and (c). The signals are not identical, but differ to some extent. The wave packet
in the chromosphere can significantly diverge from its
photospheric counterpart. This is common in the analysed sunspots. For both
techniques (lambda-meter and tracking Stokes-V zero crossing,
Figure~\ref{fig:spot-si-he-lme-vmp}) the time delay between the photosphere and
chromosphere varies, but it does not exceed 100\,s. The variations can be caused
by several factors, including the presence of the fine structure in the umbra, or
because we used different parts of the spectral line profile in two techniques.
The longest time delay between the photosphere and chromosphere in
Figure~\ref{fig:spot-si-he-lme-vmp} is 64\,s. An example for series No.\,17 in
Figure\,\ref{fig:spot-si-he-vmp-good}\,(a) shows a negative time delay of
$-$20\,s. The 1000\,km height difference and the 64\,--\,110\,s lag (see
Table\,\ref{tbl:spots}, series No. 15\,--\,18) correspond to the phase speed of
9\,--\,15 {$\mathrm {km\,s^{-1}}$}.

\par
\begin{figure} 
\centerline{
\includegraphics[width=8cm]{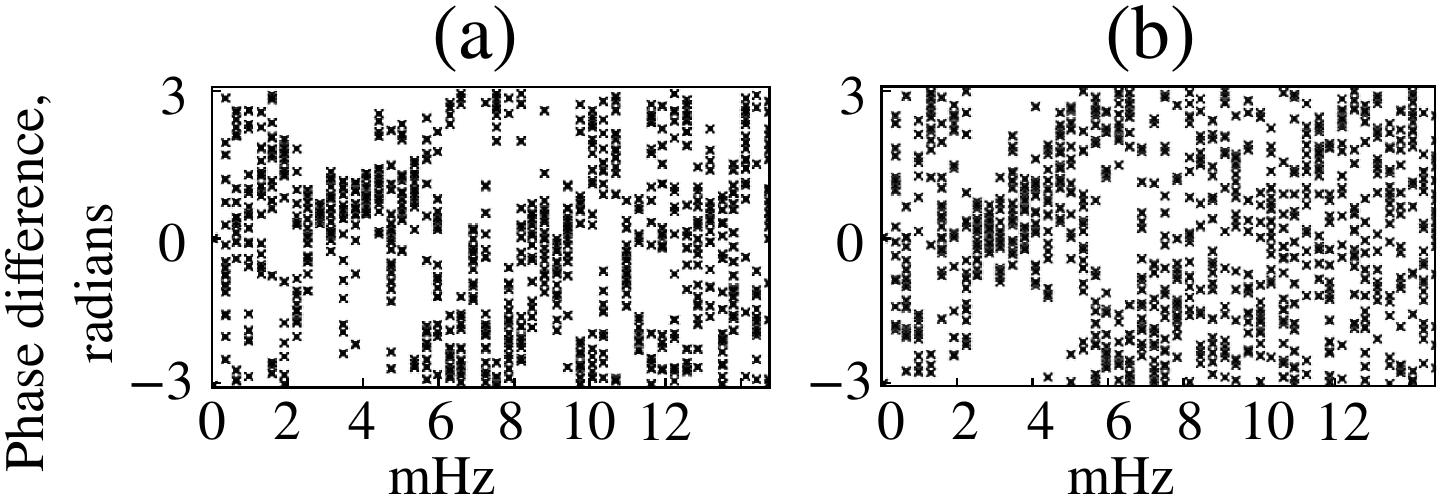}
}
\caption{Time series No.\,16.
The FFT phase difference of LOS-velocity oscillations
($V_{Si\,{I}} - V_{He\,{I}}$)
for each umbra element and specific frequency. LOS velocity
derived from (a) lambda-meter technique; (b) Stokes V-zero crossing position.
        }
\label{fig:spot-phase}
\end{figure}

 The LOS-velocity oscillation phase difference ($V_{Si\,{I}} - V_{He\,{I}}$)
for each umbra element along the entrance slit and for each specific frequency shows
weak indications of propagating oscillations in the 5\,--\,6 mHz band
(Figure\,\ref{fig:spot-phase}, time series No.\,16). The phase delays between
the photospheric and chromospheric oscillations of these frequencies deduced
from these plots are (a) 50\,s and (b) 80\,s.


\begin{figure}
\centerline{
\includegraphics[width=10cm]{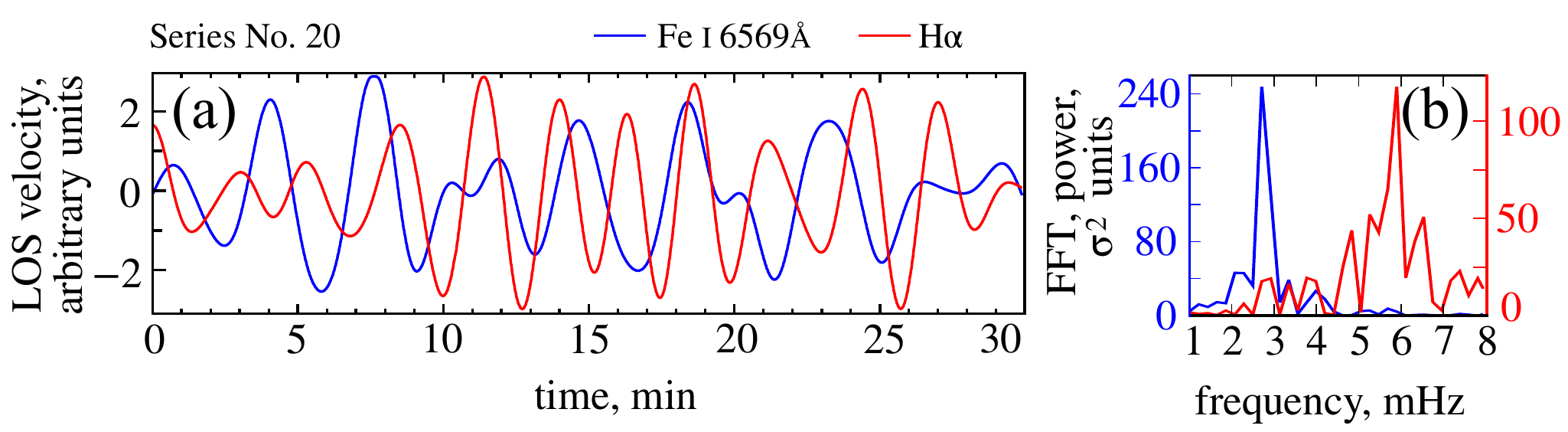}
}
\caption{Time series No.\,20, umbral region of 1$''$. (a) LOS velocity
(lambda-meter technique) filtered in the 4.6\,--\,5.6\,mHz band. (b) The
LOS-velocity power spectra at two atmospheric levels. Photospheric signal
Fe\,{\sc i} 6569\,\AA\ -- blue line; chromospheric signal H$\alpha$ 6563\,\AA\
-- red line.}

\label{fig:spot-ha-fe}
\end{figure}

\par At the Fe\,{\sc i} 6569\,\AA\,--\,H$\alpha$ line heights, the
correspondence between the photosphere and the chromosphere is less prominent
than for Si\,{\sc i} 10\,827\,\AA\,--\,He\,{\sc i} 10\,830\,{\AA}. One can
hardly find a correspondence between the two signals in
Figure\,\ref{fig:spot-ha-fe}. The power of the photospheric signal of the three-minute
band is slightly higher than the 3$\sigma^2$ level. The same is true for the other
time series obtained in the Fe\,{\sc i} 6569\,\AA\,--\,H$\alpha$ lines.

\begin{figure}
\centerline{
\includegraphics[width=13.5cm]{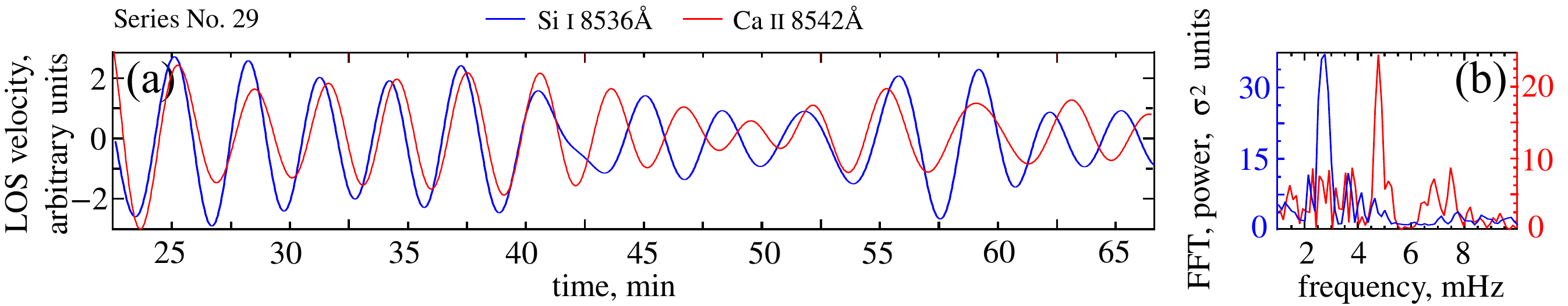}
}
\caption{Time series No. 29: (a) filtered photospheric and chromospheric
LOS-velocity signals (4.5\,--\,5.5 mHz). (b) The corresponding LOS-velocity
power spectra for two atmospheric levels.
Photosphere Si\,{\sc i} 8536\,\AA\ -- blue line;
chromosphere -- Ca\,{\sc ii} 8542\,\AA\ -- red line.
        }
\label{fig:spot-si-ca}
\end{figure}

\par For series obtained in Si\,{\sc i} 8536\,\AA\,--\,Ca\,{\sc ii} 8542\,\AA\ a
similar ambiguity between the photospheric and chromospheric heights was found,
except for one series discussed below, where it was possible to determine a time
delay. The power spectra for the chromosphere and photosphere in
Figure\,\ref{fig:spot-si-ca}~(b) are very different. Nevertheless, both signals
have a significant power within the 4.5\,--\,5.5 mHz and 7\,--\,8~mHz bands, and
there ia a wave train correspondence (Figure~\ref{fig:spot-si-ca}~(a)). The
lag between the filtered signals (4.5\,--\,5.5 mHz) in
Figure~\ref{fig:spot-si-ca}~(a) is 21\,s. This corresponds to the
upward-propagating wave at a speed of 43 {$\mathrm {km\,s^{-1}}$}, if the height
difference is 900\,km. The other series (Table\,\ref{tbl:spots}, series No.~30,
31) do not reveal any correspondence between wave trains at the two levels.

\par
\begin{figure}
\centerline{
\includegraphics[width=13.5cm]{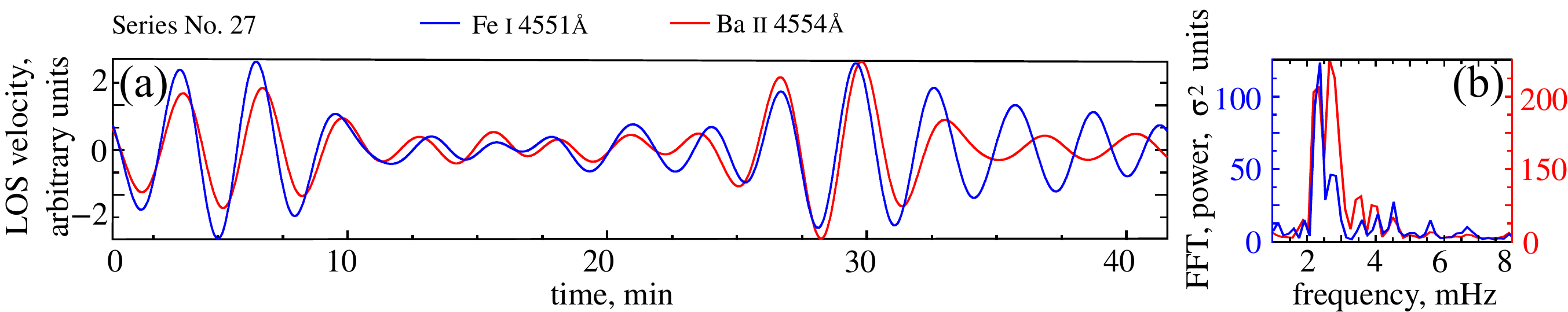}
}
\caption{Time series No.\,27, umbral region of 1$''$. (a) LOS velocities
filtered in the 5\,--\,6\,mHz band. (b) The LOS-velocity power spectra at
two atmospheric levels. Photosphere Fe\,{\sc i} 4551\,\AA\ -- blue line;
temperature minimum Ba\,{\sc ii} 4554\,\AA\ -- red line.
        }
\label{fig:spot-fe-ba}
\end{figure}

 The situation is totally different for the time series obtained in
Fe\,{\sc i} 4551\,\AA\ and Ba\,{\sc ii} 4554\,\AA\ (formation heights are 140
and 640\,km, respectively). As for the faculae, there is almost a perfect
correspondence between wave trains at two heights for the most sunspot
umbra slit positions (Figure~\ref{fig:spot-fe-ba}). For the signals shown
in Figure~\ref{fig:spot-fe-ba} the lag is about 10\,s. With a 500~km height
difference, this lag corresponds to a phase speed of 50 {$\mathrm {km\,s^{-1}}$}.

\section{Discussion and Conclusions} 
  \label{S-Conclusion}
\par We analysed LOS-velocity oscillations at different pairs of solar
atmosphere levels for faculae and sunspots. It is rare that the wave trains
correspond directly for sunspots. When the correspondence was obvious,
we measured a time delay that appeared to be short on average. The short delay
we found may indicate that the phase speed is significantly higher than the
commonly used value 4\,--\,6 {$\mathrm {km\,s^{-1}}$} for the photosphere. The
signal produced with the lambda-meter technique may be affected by the
non-magnetic elements, whereas the signal yielded by tracking the null of the
Stokes-V profile is formed only by the part where the magnetic field is present.
Obviously, this difference is very small for the sunspot umbra, where the magnetic
filling factor is close to~1. For two levels of the solar atmosphere that are
close in height ($\Delta h=$500~km, Fe\,{\sc i} 4551\,\AA\ and Ba\,{\sc ii}
4554\,\AA) the signals agree much more often than for the other analysed levels
where the height difference is significantly
higher. However, the measured time delays also correspond to the high propagation
velocities. An ambiguous results for the Fe\,{\sc i}
6569\,\AA\,--\,H$\alpha$ 6563\,\AA\ pair is likely due to the fact that at a
moderate spatial resolution of 1.5$''$ the three-minute oscillations at the Fe\,{\sc i}
6569\,\AA\ formation height are extremely weak. It is possible that a high
spatial resolution will allow one to detect domains with more powerful
oscillations at that level.

\par The interpretation of the lags and phase speeds critically depends on
the assumed line formation heights. Most of them are derived from quiet-Sun
conditions and are not valid in magnetic regions, either in faculae or
sunspots. To compare our results with those from other authors, we intentionally
used the same height difference values as they did. We can assume that
the height scale is significantly reduced in the presence of the magnetic field.
In this case, the similarity between the oscillations detected in different
lines is to be like the one revealed in the Fe\,{\sc i} 4551\,\AA\ and Ba\,{\sc
ii} 4554\,\AA\ pair. However, the analysis performed for different pairs showed
such a similarity neither for the sunspots nor for the faculae under
investigation.

\par We assume that within the observed cavity standing and propagating
waves co-exist, therefore the measured phase lag depends on the contribution of each
of the components. Their rate is determined by the resonator boundary
penetrability degree, which in real conditions changes following local
changes in temperature, pressure, and magnetic field. The ``chevrons''
on space--time diagrams clearly visible above spot umbra in the 5.5\,--\,7~mHz
band \citep{kobanov2006} may be considered as evidence of the resonator's upper
boundary transparency. The short delay raises the question of the oscillations
origin. Our results support the hypothesis of a
chromospheric resonator in sunspots \citep{zhugzhda1985,botha2011}, rather than
the hypothesis of linear wave propagation with sound speed from the photosphere
to the chromosphere \citep{centeno09}.

\par In the case of faculae the situation is different. For most of the
faculae, the five-minute oscillation periods are dominant either in the photosphere or
in the chromosphere. An ambiguous behaviour of the time lags can be explained in
the following way. Inclined magnetic flux tubes that contribute to the signal
for the lower height levels can move beyond the aperture range at the higher
levels, and \textit{vice-versa}, other flux tubes can fall into the aperture. We
discussed this possibility in \cite{kobanov2011ARep}. It would be useful
to analyse velocity signals obtained by both techniques (Stokes-V zero-crossing
and lambda-meter). The magnetic filling factors are significantly less than
one in the facular regions at the photospheric level \citep{mpil1997,mug1992}.
This would lead to different results obtained by both techniques for the
photosphere, while in the chromosphere the expansion of the flux tubes might
smooth this difference (see Figure 3, \cite{khomenko2008ApJ}). However,
these measurements allow one to determine the non-magnetised component
contribution (\textit{e.g}. surrounding p modes). In addition, the rare
correspondence between the signals at two heights may be explained by the
dispersive properties of the medium that smear the wave packet significantly.

\par Observations made simultaneously for many lines formed at
different heights would certainly clarify the problem. The spectral line set
presented here seems quite appropriate for this task.

\vskip2.5cm

\small{\textbf{Acknowledgements}. This study was supported in part by the RFBR
research project No.: 12\hbox{-}02\hbox{-}33110 mol\_a\_ved \, and \,
the Grant of the President of the Russian Federation
No.: MK\hbox{-}497.2012.2, Russian Federation
Ministry of Education and Science state contract No.: 14.518.11.7047 and
agreement No.: 8407.}

\bibliographystyle{spr-mp-sola}
\bibliography{kolobov}

\end{document}